\documentclass[prl,twocolumn,preprintnumbers,amsmath,amssymb]{revtex4}
\usepackage{color}
\usepackage{hyperref}
\usepackage{graphicx}
\usepackage{float}
\usepackage{epstopdf}
\usepackage{multirow}

\begin{document}
\title{Towards Fully Converged GW Calculations for Large Systems}

\author{Weiwei Gao} 
\author{Weiyi Xia}
\author{Xiang Gao}
\author{Peihong Zhang}

\affiliation{Department of Physics, University at Buffalo,
State University of New York, Buffalo, New York 14260, USA}
\date{\today}

\begin{abstract}
Although the GW approximation is recognized as one of the most accurate theories
for predicting materials excited states properties, scaling up conventional GW calculations 
for large systems remains a major challenge.
We present a powerful and simple-to-implement method that can drastically
accelerate fully converged GW calculations for large systems. 
We demonstrate the performance of this new method by calculating the
quasiparticle band gap of MgO supercells.
A speed-up factor of nearly two orders of magnitude is achieved  for a system contaning 256 atoms  (1024 velence electrons) 
with a negligibly small numerical error of $\pm 0.03$ eV.  
\end{abstract}
\maketitle


Accurate predictions of excited states properties are critical for computational screening and 
design of materials for energy and electronics applications.  While density functional 
theory (DFT)\cite{hohenberg1964} within the local density approximation (LDA)\cite{kohn1965}
or the generalized gradient approximation (GGA)\cite{perdew1986}
has been very successful in describing materials ground-state properties, the Kohn-Sham (KS) eigenstates
in principle cannot be compared directly with quasiparticle (QP) excitations. 
The GW approximation\cite{hedin1965,hybertsen1986, godby1988}, systematically derived from the perturbative expansion of electron self-energy, is widely recognized as one of the most accurate methods for
predicting excited states properties of materials.

Unfortunately, despite the tremendous success, {\it accurate} and {\it efficient} predictions of
excited-states properties of complex solid systems remain a major challenge due to complication of
the convergence issue~\cite{bruneval2008,berger2010,shih2010,samsonidze2011,deslippe2013} and the unfavorable 
scaling of the computational cost (both in terms of the computational time and the storage and memory requirements) 
with respect to the system size. This is particularly true for systems containing substantially localized
electronic states and/or with large unit cells (e.g., nanostructures, complex multinary compounds, surfaces and
interfaces, and defects in solids).  Although zinc oxide (ZnO) has probably received the most research
attention,\cite{shih2010,friedrich2011,stankovski2011,deslippe2013} 
the convergence problem of GW calculations is not just limited to this single
material. In fact, GW calculations for many systems, including MgO, CuCl, AgBr, and CdO
(to name a few), suffer similar convergence problems. 

The GW approximation for the electron self-energy $\Sigma(\mathbf{r},\mathbf{r}';E)$ is
\begin{equation}
\Sigma=i\int\frac{dE'}{2\pi}e^{-i\delta E'}G(\mathbf{r},\mathbf{r}';E-E')W(\mathbf{r},\mathbf{r}';E'),
\end{equation}
where $\delta=0^+$, $G$  is the one-particle Green's function, and $W$ is the screened Coulomb
interaction, which is related to the dielectric function $\epsilon$ and the bare Coulomb interaction $v$ as $W=\epsilon^{-1}v$.
In conventional first-principles GW methods\cite{hybertsen1986, godby1988}, both the Green's function and
the dielectric functions are constructed using the KS eigenstates, and, in principle, all states in the Hilbert space
of the KS Hamiltonian should be included in the summations. 
Within the random phase approximation (RPA), the dielectric function is related to the electron polarizability $\chi^0$  via
$\epsilon=1-v\chi^0$, where
\begin{align}
\label{eq:chi0}
 &\chi^0_{\mathbf{G,G}'}(\mathbf{q},\omega) =
\frac{1}{2}\sum_{c,v,\mathbf{k}} M_{vc}(\mathbf{k,q,G}) M^*_{vc}(\mathbf{k,q,G'}) \times \nonumber \\
&\bigg( \frac{1}{ E_{v\mathbf{k+q}} - E_{c \mathbf{k}} - \omega + i\delta}  + 
\frac{1}{ E_{v\mathbf{k+q}} - E_{c \mathbf{k}} + \omega + i\delta} \bigg),
\end{align}
where $M_{vc}(\mathbf{k,q,G})$=$\langle v,\mathbf{k+q}| e^{i(\mathbf{q+G})} | c,\mathbf{k}\rangle$,  and $v$ and $c$
index valence and conduction bands.
The calculation of the COH self-energy also involves a summation over conduction bands:~\cite{hybertsen1986,deslippe2012}
\begin{align}
\label{eq:sigmach}
&\langle n \mathbf{k} | \Sigma_\text{COH}(E) | n' \mathbf{k} \rangle = \nonumber \\
&\frac{i}{2\pi}\sum^{\substack{\text{all}\\ \text{bands}}}_{n''}\sum_{\mathbf{qGG'}} M^*_{n'' n}(\mathbf{k,-q,-G}) M_{n'' n'}(\mathbf{k,-q,-G'}) \times \nonumber \\
& \int\limits^\infty_0 dE' \frac{[\epsilon^r_{\mathbf{G,G}'}(\mathbf{q};E')]^{-1} - [\epsilon^a_{\mathbf{G,G}'}(\mathbf{q};E')]^{-1}}{E-E_{n''\mathbf{k-q}}-E'+i\delta}v(\mathbf{q+G}'). 
\end{align}
where $\epsilon^{r}$ ($\epsilon^{a}$) is the retarded (advanced) dielectric function.

The band summations in the calculations of dielectric function and COH self-energy should in principle include all conduction (empty) states in the Hilbert space of the KS Hamiltonian.
In practical calculations, truncations are almost always applied. Since the convergence behavior of GW calculations 
could be very different from materials to materials, it is often difficult to predict {\it a priori} proper truncation parameters to ensure
fully converged GW results. For large systems, it is extremely difficult
(if possible at all) to perform fully converged GW calculations within this scheme.
In this letter, we present an effective method which promises to speed up
fully converged GW calculations by up to two orders of magnitude for large systems containing hundreds of atoms and is also
easy to implement within available GW packages.

{\it Computational details.}--- 
We use MgO as an example to illustrate the difficulty of scaling up conventional GW calculations for large systems and demonstrate
the performance of the newly developed method.
All GW results presented in this letter are calculated within the so-called G$^0$W$^0$ approach using
a modified version of the \texttt{BerkeleyGW}\cite{deslippe2012} package.
The Hybersten-Louie generalized plasmon-pole (HL-GPP) model\cite{hybertsen1986} is used to extend
the static dielectric function to finite frequencies.
The new method presented in this letter, however, can be applied to other GW implementations including
various self-consistent GW approaches.
The DFT calculations within the LDA are carried out using the \texttt{PARATEC}\cite{paratec} code.
Norm-conserving Troullier-Martin pseudopotentials\cite{troullier1991} are used.
The kinetic energy cutoff for the plane-wave expansion of the KS wave functions is set at 80 Ry.

\begin{figure}[t]
  \centering
  \includegraphics[width=0.48\textwidth]{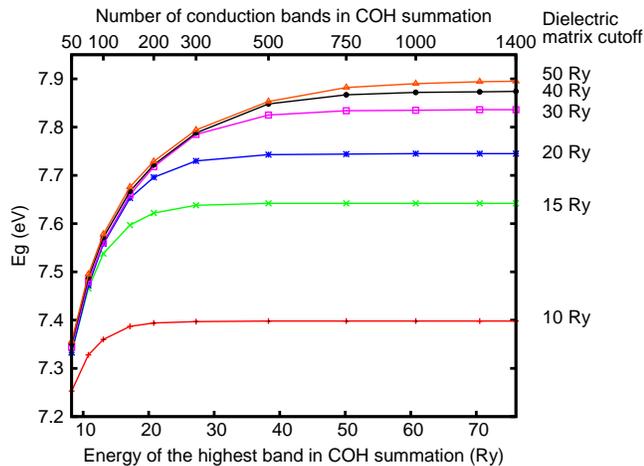}
  \caption{(Color online) Calculated quasiparticle band gap of MgO as a function of the number of conduction bands $N_\text{c}$,
or equivalently, the energy of the highest conduction band,
included in the COH self-energy calculations and the kinetic energy cutoff for the
dielectric matrices.}
  \label{fig:MgOconvergence}
\end{figure}

Figure~\ref{fig:MgOconvergence} shows the convergence behavior
the calculated quasiparticle band gap of MgO (two-atom primitive cell) as a function of the number of conduction bands included in the 
COH self energy summation and the kinetic energy cutoff for the dielectric matrices. 
Depending on various cutoff parameters such as the number 
of bands included in the Coulomb-hole self-energy summation and the kinetic cutoff for the dielectric matrices,
the calculated QP band gap varies from 7.25 to 7.90 eV.
Once fully converged, we obtain a QP band gap of about 7.90 eV, in good agreement with the experimental value
of 7.78 eV.\cite{whited1969}
As it is shown in Fig~\ref{fig:MgOconvergence}, a high kinetic cutoff 
($|\mathbf{G}_{\mathrm{cut}}|^2/2\sim$ 40 Ry) for the dielectric matrices 
$\epsilon_{\mathbf{G},\mathbf{G}'}(\mathbf{q},\omega)$ and a large number 
of conduction bands ($N_c\sim$ 800) in the COH summation are required. 
If a small kinetic energy cutoff for the dielectric matrices is used, the band gap converges quickly but  prematurely with respect to
the number of band included in the summation. 

With today's massively parallel computers,
fully converged GW calculations for small systems such as 2-atom MgO can be done easily.
The problem, however, quickly becomes intractable when one attempts to scale up the calculations 
for large systems containing hundreds of atoms because
the number of conduction bands required in the COH and dielectric function
calculations scales linearly with the system size (i.e., number of atoms). 
To put this problem in perspectives, suppose now we would like to
carry out GW calculations for a 64-atom MgO supercell containing a defect (e.g., an F-center),
the required number of bands would increase by 32 (64/2) times.
Not only it is difficult to generate so many wave functions, storing these wave functions 
would also pose a serious challenge, not to mention performing the band summations in the GW calculations.

\begin{figure}[h]
   \centering
    \includegraphics[width=0.48\textwidth]{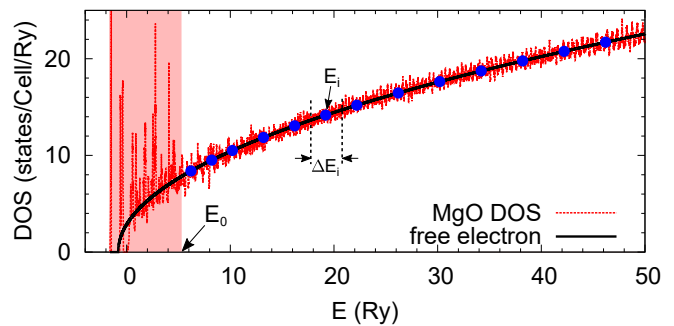}
\caption{ \label{fig:mgodos} (Color online) Comparison between the DOS of MgO
calculated within the LDA (red) and that of free electron gas (black).
The DOS for the free electron gas is shifted with respect to the averaged 
exchange-correlation potential of MgO.}
\end{figure}

Our approach is inspired by the observation that the density of states (DOS)
of most materials can be divided into two regions: a low energy
region in which the behavior of electronic states is strongly
affected by the crystal potential and a high 
energy region in which kinetic energy dominates and the DOS scales as 
$\sqrt{E-V_{xc}(0)}$ as that for free electron systems, where $V_{xc}(0)$ is 
the averaged exchange-correlation potential of the material.
Figure~\ref{fig:mgodos} compares the DOS of MgO calculated within the LDA and that of 
the free-electron gas scaled with the volume of the unit cell of MgOl.
The DOS below certain energy ($E_0$ shown in Fig.\ref{fig:mgodos}) is what distinguishes
one material from another. Above $E_0$, the DOS can be well-approximated by
that of free-electron gas.

This observation leads us to propose that the band summation in the GW calculations
be divided into two regions: a low energy region (shaded area in Fig.\ref{fig:mgodos}) in which the band summation is
carried out explicitly and a high-energy region in which the band summation is
replaced (approximated) by an energy integration. For example, 
the static polarizability $\chi^0$ is rearranged in the following form:
\begin{align}
\label{eq:chi02}
&\chi^0_{\mathbf{G,G}'}(\mathbf{q},\omega=0) \approx \nonumber \sum^{N_0}_{c} \Bigg[
\sum_{v\mathbf{k}}
\frac{M_{vc}(\mathbf{k,q,G}) M^*_{vc}(\mathbf{k,q,G'}) }{
E_{v\mathbf{k+q}} - E_{c \mathbf{k}}} \Bigg]\nonumber \\
& + \int\limits_{E_0} \sum_{v\mathbf{k}} \frac{M_{vE'}(\mathbf{k,q,G})
M^*_{vE'}(\mathbf{k,q,G'}) }{ E_{v\mathbf{k+q}} - E'} g(E') \text{d}E',
\end{align}
where $g(E)=\frac{\Omega}{\pi^2}\sqrt{2(E-V_{xc}(0))}$ is the approximated
DOS for the system with a cell volume $\Omega$, $E_0$ corresponds
to the energy of band $N_0$, and $M_{vE}(\mathbf{k,q,G})$ 
are the matrix elements between the valance band $v$ and the conduction band at
(or near) the energy $E$.  
In practical calculations, we find that the number of bands
$N_0$ can be limited to less than 10 per atom for most systems.
The auxiliary integral part can be carried out using simple numerical integration techniques, for example,
\begin{equation}
\sum^{N_E}_{i=1} \sum_{v\mathbf{k}} \frac{M_{vE_i}(\mathbf{k,q,G})
M^*_{vE_i}(\mathbf{k,q,G'}) }{ E_{v\mathbf{k+q}} - E_i} g(E_i)\Delta E_i,
\label{eq:energyintegration}
\end{equation}
where $N_E$ is the number of sampling points used in the numerical integration as shown schematically in Fig.\ref{fig:mgodos} with large blue dots.
The same approach is applicable to the calculation of the COH self-energy (Eq.\ref{eq:sigmach})..
The integration can be carried on a uniform or nonuniform energy grid, and a typical energy 
step $\Delta E$ can range from 1 to 5 eV. Higher order numerical integration techniques may also be applied.

\begin{table}[h]
\caption{Calculated GW band gap (in eV) of a 16-atom MgO cell: accuracy and speed-up factor of the conduction band integration method. }
\begin{tabular}{c | c | c |  c | c |c |rc} 
\hline\hline
\multirow{2}{*}{\parbox{0.7 cm}{$\Delta E$ (eV) }}  &  \multicolumn{2}{c|}{\parbox{2.2 cm}{New method }}&
\multicolumn{2}{c|}{\parbox{2.2 cm}{Conventional method}} &\multirow{2}{*}{\parbox{1.4 cm}{ Speed-up factor}}   & \multirow{2}{*}{$\Delta E_g$} \\ 
\cline{2-5}
 &{\parbox{1.25 cm}{$N_0+N_E$}} &  {\parbox{1.2 cm}{$E_g$}}                 &  {\parbox{1.2 cm}{ $N_c$ }}& $E_g$                                    &                              &\\ \hline

1.5 & 895 & 7.84 & \multirow{6}{*}{ 8,000}& \multirow{6}{*}{7.86}&8.9& $-0.02$ \\
2.0 & 640 & 7.86 &&&12.5& $0.00$\\
2.5 & 510 & 7.88 &&&15.7& $+0.02$\\
3.0 & 435 & 7.88 &&&18.4& $+0.02$\\
3.5 & 385 & 7.86 &&&20.8& $0.00$\\
4.0 & 320 & 7.84 &&&25.0& $-0.02$\\
\hline\hline
\end{tabular}
\label{table:16atom}
\end{table}

We now examine the stability and performance of this new method using a
reasonably small system, an MgO supercell containing 16 atoms, as an example.
For this system, we choose $N_0$ to be 128. In other words, 8 conduction bands per atom are included in
the explicit summations in the calculation of the dielectric matrices and the COH self-energy. 
The auxiliary integral part is calculated on a uniform energy grid with a step $\Delta E$ ranging from 1.5 to 4.0 eV, and the
integration is carried out up to an energy which is equivalent to including 8,000 (or 1,000 per 2-atom cell) conduction bands. 
The kinetic energy cutoff for the dielectric matrix is set at 40 Ry.

For such a small system, we can compare directly the
results calculated using the new method and the conventional band-by-band summation approach,
as shown in Table~\ref{table:16atom}. As the energy integration step varies
from 1.5 to 4 eV (while keeping $N_0$ fixed at 128), the number of conduction bands plus
the number of the energy integration points decreases from 895 to 320, to be compared with 8,000 required in conventional calculations
to achieve the same level of convergence, representing a speed-up factor
of 9 to 25. 
It is rather encouraging that the results calculated using this new method are  
insensitive to the energy integration step $\Delta E$, showing the stability of the this method.
The GW band gap calculated with the conventional method is 7.86 eV, and that
obtained from the new method only deviates by a negligible small amount of $\pm 0.02$ eV.

Our method reduces the computational time of fully converged GW calculations for a 16-atom MgO 
supercell from about 5 hours to 15 minutes using 64 computing cores (8 Intel Xeon E5-2650 processors).
We would like to mention that within this new approach, one only needs to calculate (and store) 
the wave functions of the $N_0$ low energy states plus selected  high-energy conduction states at or near 
predefined energy grid points.  The folded-spectrum method
developed by Wang and Zunger~\cite{wang1994} and an alternative
approach proposed by Tackett and Ventra~\cite{tackett2002} are
ideal for this purpose. Using these methods, one can efficiently
calculate KS eigenfunctions at or near specified energy grid points without the need
to calculate all KS states. Therefore, the speed-up factor shown in 
Table~\ref{table:16atom} also indicates the reduction in the required memory and disk space
associated with storing the wave functions in GW calculations. 

The most interesting and important aspect of our method is that the speed-up factor
actually increases with increasing system size, enabling fully converged GW calculations for
large systems containing hundreds of atoms. This is because with a given 
integration step size $\Delta E_i$, the weight of a single integration point, $g(E_i)\Delta E_i$ in Eq.\ref{eq:energyintegration}, which
measures effectively the number of states contributing to the summation represented by a single state at $E_i$, scales
linearly with the system size. To demonstrate that our method can indeed be scaled up for
large systems, we have carried out GW calculations for MgO systems containing 2 to 256 atoms.

\begin{table}[h]
\caption{Performance of the new method with increasing system size. }
\begin{tabular}{c | c | c |  c | c |c |rc} 
\hline\hline
\multirow{2}{*}{\parbox{0.9 cm}{\# of atoms}}  &  \multicolumn{2}{c|}{\parbox{2.2 cm}{New method }}&
\multicolumn{2}{c|}{\parbox{2.2 cm}{Conventional method}} &\multirow{2}{*}{\parbox{1.4 cm}{ Speed-up factor}}   & \multirow{2}{*}{$\Delta E_g$} \\ 
\cline{2-5}
 &{\parbox{1.25 cm}{$N_0+N_E$}} &  {\parbox{1.2 cm}{$E_g$}}                 &  {\parbox{1.2 cm}{ $N_c$ }}& $E_g$                                    &                              &\\ \hline

2 & 170& 7.86 & 1,000& \multirow{6}{*}{7.86}&5.9& $0.00$ \\
16& 320& 7.84 &8,000&& $25.0$&$-0.02$\\
64& 920& 7.89 &32,000&&34.8& $+0.03$\\
128&1060 & 7.83 &64,000&&60.4& $-0.03$\\
256& 1580 & 7.86 &128,000&&81.0& $0.00$\\
\hline\hline
\end{tabular}
\label{table:gwall}
\end{table}

Table~\ref{table:gwall} shows the performance of our method, both in terms of accuracy and
the speed-up factor. A speed up factor of over 80 times is achieved for the largest system containing
256 atoms (1,024 valence electrons), and the numerical integration error, measured by the calculated band gap, is only $\pm 0.03$ eV
for all systems. Thus the greater speed-up for larger systems does not compromise the accuracy of this method,
and there is still room for improvement (for example, with improved numerical integration techniques).
We would like to mention that such fully converged GW calculations for large oxide systems containing over 200 atoms would be
nearly impossible using the conventional approach due to the vast amount of computational resources
required (both in terms of the computational time and the memory and storage requirements).

Table~\ref{table:dielectric} compares the calculated  macroscopic dielectric constant $\epsilon_{\infty}=\lim_{\mathbf{q\rightarrow 0}}[1/\epsilon^{-1}_{0,0}(\mathbf{q},0)]$
using the conventional and the new approach. Only the result for the 16-atom system is calculated using the conventional GW approach,
which includes 8,000 conduction bands in the summation.
As it is shown in the Tables \ref{table:gwall} and \ref{table:dielectric}, our method can not only reproduce accurately the GW band gap but also other quantities such as the dielectric constant at a fraction of the computational
cost compared with the conventional approach.

\begin{table}[h]
\centering
\caption{Comparison between the calculated macroscopic dielectric constant of MgO using the conventional (for the 16-atom system) and new methods (for systems with 64, 128 and 256 atoms).}
\begin{tabular} {c |  p{0.08\textwidth}   p{0.08\textwidth}   p{0.08\textwidth}   p{0.05\textwidth}  }
\hline\hline
\# of atoms            & 16 &  64 & 128 & 256\\
\hline
$\epsilon_{\infty}$ &3.012&3.010&3.012&3.011\\
\hline\hline
\end{tabular}
\label{table:dielectric}
\end{table}

Considering the importance of predicting materials excited states properties,
it is not surprising that there has been much work on improving the the efficiency of the GW formalism for large 
systems beyond simple crystals.
One direction is to introduce better (optimized) bases  for representing the dielectric matrix and/or the self-energy operator
beyond plane waves and KS eigen states.~\cite{umari2009,caruso2013,pham2013}.
The second direction is to reduce the number of conduction bands $N_c$ by replacing the summation over high-energy 
states with computationally tractable terms~\cite{bruneval2008,berger2010,tiago2006,kang2010,deslippe2013},
or completely removing the necessity of summing over unoccupied states~\cite{giustino2010,lambert2013,govoni2015}.

These efforts underline the urgency of developing efficient GW methods to meet the
challenge of fast and accurate predictions of the excited states properties of large systems such as
nanostructures, defects in solids, and surfaces and interfaces.
Some of these approaches, however,  may require substantial modifications (or completely rewriting) existing GW codes, and the
study of the convergence behavior of these methods is still in the early stage.
Our approach is conceptually simple and is very easy to be implemented in well-developed GW packages.
More importantly, we have demonstrated affordable and fully converged GW calculations for large and {\it hard-to-converge} systems such as MgO.

In summary, we present an efficient integration method that drastically
speeds up fully converged GW quasiparticle calculations for 
large systems. Our method takes advantage of the fact
that the DOS of most materials resembles that of the free electron gas at high energies, thus the
summation over high-energy conduction bands in both the dielectric function and
self-energy calculations within the GW approximation can be well approximated by a numerical integration
on a sparse energy grid. Using this new method, we can now carry out fully
converged GW calculations for large systems without the need to perform
tedious and wasteful convergence tests. We have demonstrated a nearly two orders of magnitude speed-up
of GW calculations for a 256-atom MgO model system while achieving a numerical accuracy of $\pm 0.03$ eV for the predicted band gap.
Although we have presented results calculated within the so-called G$^0$W$^0$ approximation and using the
HL-GPP model for the dielectric function, our approach can be applied to various levels of self-consistent
GW calculations and/or GW calculations without the use of plasmon-pole models.

\section{acknowledgement}
We thank Steven G. Louie for his helpful discussions.
This work is supported by the NSF under Grant No.
DMR-0946404 and DMR-1506669, and by the SUNY
Networks of Excellence. We acknowledge the computational support provided by the Center
for Computational Research at the University at Buffalo, SUNY.

\bibliographystyle{apsrev}
\bibliography{GW}
\end{document}